\documentclass{iaus}
\usepackage{graphicx}

\title[VO-KOREL] %% give here short title %%
{VO-KOREL: A Fourier disentangling  service of Virtual Observatory}

\author[Petr \v{S}koda, Petr Hadrava \& Jan Fuchs]  %% give here short author list%%
{Petr \v{S}koda,
Petr Hadrava
\and Jan Fuchs}

\affiliation{Astronomical Institute, Academy of Sciences, \\
Fri\v{c}ova 298,
CZ-251\,65 Ond\v{r}ejov, Czech Republic \\
email: {\tt skoda@sunstel.asu.cas.cz} \\[\affilskip]
}
\pubyear{2011}
\volume{282}  %% insert here IAU Symposium No.
\pagerange{119--126}
\jname{From Interacting Binaries to Exoplanets: Essential Modeling Tools}
\editors{A.C. Editor, B.D. Editor \& C.E. Editor, eds.}
\begin{document}

\maketitle

\begin{abstract}

VO-KOREL is a  web service exploiting the technology of Virtual Observatory for
providing the astronomers with the intuitive graphical front-end and distributed
computing back-end running the most recent version of Fourier disentangling
code KOREL.

The system integrates the ideas of the  e-shop basket, conserving the privacy
of every user by transfer encryption and access authentication, with  features
of laboratory notebook, allowing the easy housekeeping of both input parameters
and final results,  as well as it explores a newly emerging technology of cloud
computing.

While the web-based front-end allows the user to submit data and parameter
files, edit parameters, manage a job list, resubmit or cancel running jobs and
mainly watching the text and graphical results of a disentangling process, the
main part of the back-end  is a simple job queue submission system executing in
parallel multiple instances of  FORTRAN code KOREL. This  may be easily
extended for GRID-based deployment on massively parallel computing clusters.

The short introduction into underlying technologies is given, briefly
mentioning advantages as well as  bottlenecks of the  design used. 

\end{abstract}
\firstsection % if your document starts with a section,
\section{Virtual Observatory}
The Virtual observatory (hereafter VO) is a global infrastructure of
distributed astronomical archives and data processing services enabling the
standardised discovery and access to the astronomical data worldwide as well as
a large set of powerful tools for scientific  analysis and visualisation 
(\cite[Solano 2006]{solano}).
VO supports mainly the  multi-wavelength research
or discovery of rare
objects by cross-matching huge catalogues.
% or
%complex processing and analysis of vast amount of astronomical data on a
%distributed (super)computers.  

For interaction with the user as well as other computers the VO uses the modern
technology of  Web Services (WS).  The WS is typically complex processing
application using the web technology  to transfer input data  to the main
processing back-end and the results (after intensive number crunching) back to
user.  All this can be done using only an ordinary web browser (and in
principle the science may be done on the fast palmtop or advanced mobile
phone).  An example of WS is the e-shop or ticket reservation system.

\section{Universal worker service} 
The Universal Worker Service --- UWS (\cite[Harrison \& Rixon 2010]{UWS}) is
one of many standards of VO describing the exact operation patterns of WS
supporting the execution of  multiple jobs in an asynchronous way, allowing the
simple control of jobs (e.g. changing their execution limits, timeouts and
expiration) as well as easy access to the results using the web technology.
The current version of UWS  is based on a newly re-discovered RESTful
technology ({\cite[Fielding 2000]{Fielding}).

The technology of WS allows to provide the complicated computing code as a
service running in virtualized infrastructure of a  large commercial provider
or on a dedicated servers of the software provider instead of providing only
the source code.  This  idea has been promoted by the large IT companies under
the terms of "Software as a service" and it is part of wider business called
{\it Cloud computing} (\cite[Foster et al. 2008]{Foster}).  The advantages of
this approach from the developer's point of view are clear:

\begin{itemize}
\item The only SW needed by user is only a tiny web browser
\item There is the only one, current, well tested version of
the code (and documentation), maintained and updated often directly by its author
\item The computing of various models may be controlled even from the smart
mobile phone over the slow connection as the large data are uploaded only
once and most of the investigation requires only changing several numbers in a
parameter file and re-submission of a job directly from web browser.
\item The web technology provides the easy way of interaction (forms) and graphics output (in-line images) 
\end{itemize}

\section{VO-KOREL web service}
The principal use of VO-KOREL is similar to e-shop portal, starting with user
registration. Every set of input parameters creates a job, which may be run in
parallel with others. The user may stop or remove them, can return to the
previous versions etc.  All user communication is encrypted and the user can
see only his/her jobs.  The service may be accessed from the KOREL portal at
Astronomical Institute in Ond\v{r}ejov at address  {\tt
http://stelweb.asu.cas.cz/vo-korel}.

The VO-KOREL web service requires to upload the data file {\tt korel.dat} and
parameter file {\tt korel.par} (and optionally the template(s) for
template-restricted disentangling {\tt korel.tmp}) in the format described in
KOREL manual (last chapters in \cite[Hadrava 2009]{hadrava}).

Several sets (they may be as well compressed in form of an tar.gz file)
can be uploaded to the server (withing the disk quota allowed for the user) and
the execution of the jobs may be postponed even until next login of the user.
The user can decide about priorities which jobs to compute, as the system can
run only several jobs of every user, the other are queued until the computing
resources (memory, CPU) are available.

\section{Conclusions}
The VO-KOREL service is not only giving comfortable environment for Fourier
disentangling of spectra, but it is a test-bed of general cloud infrastructure for
execution of most scientific computationally intensive codes, like models of
stellar atmospheres or special processing of complex data sets. 
 \end{document}